\documentclass[12pt,twoside,epsfig]{article}
\usepackage{fleqn,espcrc1}

\usepackage{graphicx}
\usepackage[figuresright]{rotating}

\newcommand{\AmS}{{\protect\the\textfont2
  A\kern-.1667em\lower.5ex\hbox{M}\kern-.125emS}}

\def\bm{\boldmath}
\def\mb{\mbox}

\def\bge{\begin{equation}}
\def\ene{\end{equation}}
\def\bg{\begin{eqnarray}}
\def\en{\end{eqnarray}}

\def\vr{\vec{r}}

\title{Effect of changes in meson properties  
in a nuclear medium:\\ 
$J/\Psi$ dissociation in nuclear matter, 
and meson-nucleus bound states\footnote{ADP-00-23/T406, UGI-00-8\\
Work supported by 
the Australian Research Council, and the 
Forschungzentrum J\"{u}lich}} 

\author{K. Tsushima\address{Special Research Center for 
the Subatomic Structure of Matter 
and Department of Physics and Mathematical Physics, 
University of Adelaide, SA 5005, Australia}, 
A. Sibirtsev\address{Institut f\"ur Theoretische Physik, 
Universit\"at Giessen, D-35392 Giessen, Germany},
K. Saito\address{Physics Division, Tohoku College of Pharmacy,
Sendai 981-8558, Japan},
A. W. Thomas$^{\rm a}$, 
D.H. Lu$^{\rm a}$\footnote{Present address: Department of 
Physics, National Taiwan University, Taipei 10617, Taiwan}
}

\begin{document}

\maketitle

\begin{abstract}
We discuss the effect of changes in meson properties in a nuclear medium 
on physical observables, notably, $J/\Psi$ dissociation on pion   
and $\rho$ meson comovers in relativistic heavy ion collisions, and 
the prediction of the $\omega$-, $\eta$- and $\eta'$-nuclear bound states.
\end{abstract}

\section{Mean-field potentials for mesons and baryons in QMC}

This report is based on the quark-meson 
coupling (QMC) model~\cite{Guichon}, 
which has been successfully applied to many problems in nuclear
physics~\cite{Guichonf,Tsushimak,Tsushimaeta,Tsushimad,Tsushimaj}. 
A detailed description of the Lagrangian density and the
mean-field equations of motion are given in 
Ref.~\cite{Guichonf}.
The Dirac equations for the quarks and antiquarks in hadron bags 
($q = u,\bar{u},d$ or $\bar{d}$, hereafter, and including up to 
$s,\bar{s},c,$ and $\bar{c}$)
neglecting the Coulomb force, 
are given by
($|\mbox{\boldmath $x$}|\le$ bag 
radius)~\cite{Guichonf,Tsushimak,Tsushimaeta,Tsushimad,Tsushimaj}:
\begin{eqnarray}
\left[ i \gamma \cdot \partial_x -
(m_q - V^q_\sigma)
\mp \gamma^0
\left( V^q_\omega +
\frac{1}{2} V^q_\rho
\right) \right] 
\left( \begin{array}{c} \psi_u(x)  \\
\psi_{\bar{u}}(x) \\ \end{array} \right)
&=& 0, 
\left( (V^q_\omega -\frac{1}{2}V^q_\rho)\, {\rm for}\,
\left(\begin{array}{c} \psi_d  \\ 
\psi_{\bar{d}}\\ \end{array} \right) \right),
\label{diracu}
\\
\left[ i \gamma \cdot \partial_x - m_{s,c} \right]
\psi_{s,c} (x)\,\, ({\rm or}\,\, \psi_{\bar{s},\bar{c}}(x)) &=& 0. 
\label{diracsc}
\end{eqnarray}
The mean-field potentials for a bag in nuclear matter
are defined by $V^q_\sigma{\equiv}g^q_\sigma
\sigma$,
$V^q_\omega{\equiv}$ $g^q_\omega
\omega$ and
$V^q_\rho{\equiv}g^q_\rho b$,
with $g^q_\sigma$, $g^q_\omega$ and
$g^q_\rho$ the corresponding quark-meson coupling
constants. 

\begin{figure}[htb]
\begin{minipage}[htb]{75mm}
\vspace*{-2em}
\includegraphics[height=9cm,width=7cm]{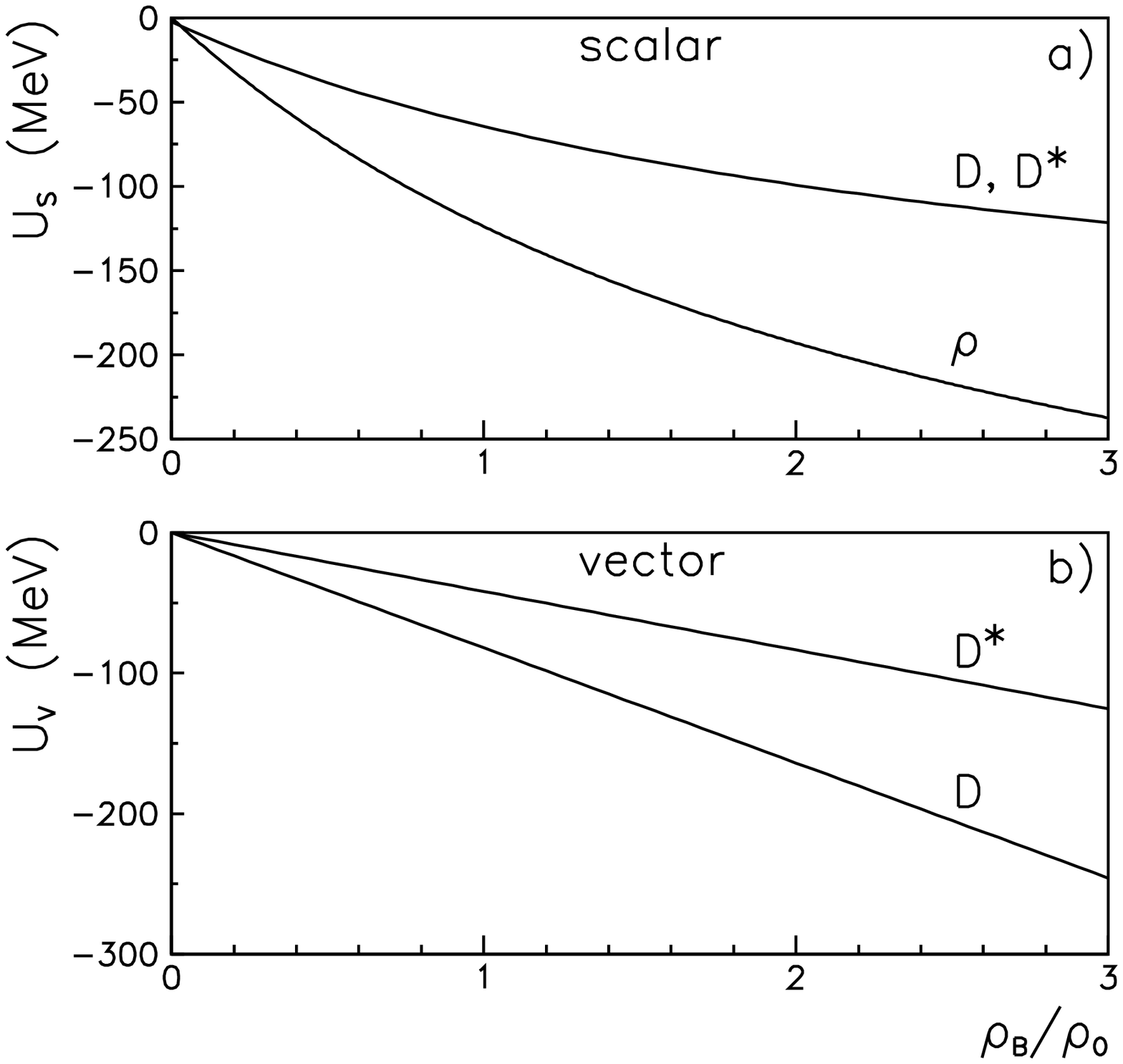}
\vspace{-3em}
\caption{
Scalar and vector potentials for 
the $D$, $D^*$ and $\rho$ mesons in symmetric nuclear matter 
($\rho_0$=0.15 fm$^{-3}$).
\vspace{-1em}}
\label{dpot}
\end{minipage}
\hspace{\fill}
\begin{minipage}[htb]{75mm}
\vspace*{-2em}
\includegraphics[height=4cm,width=6cm]{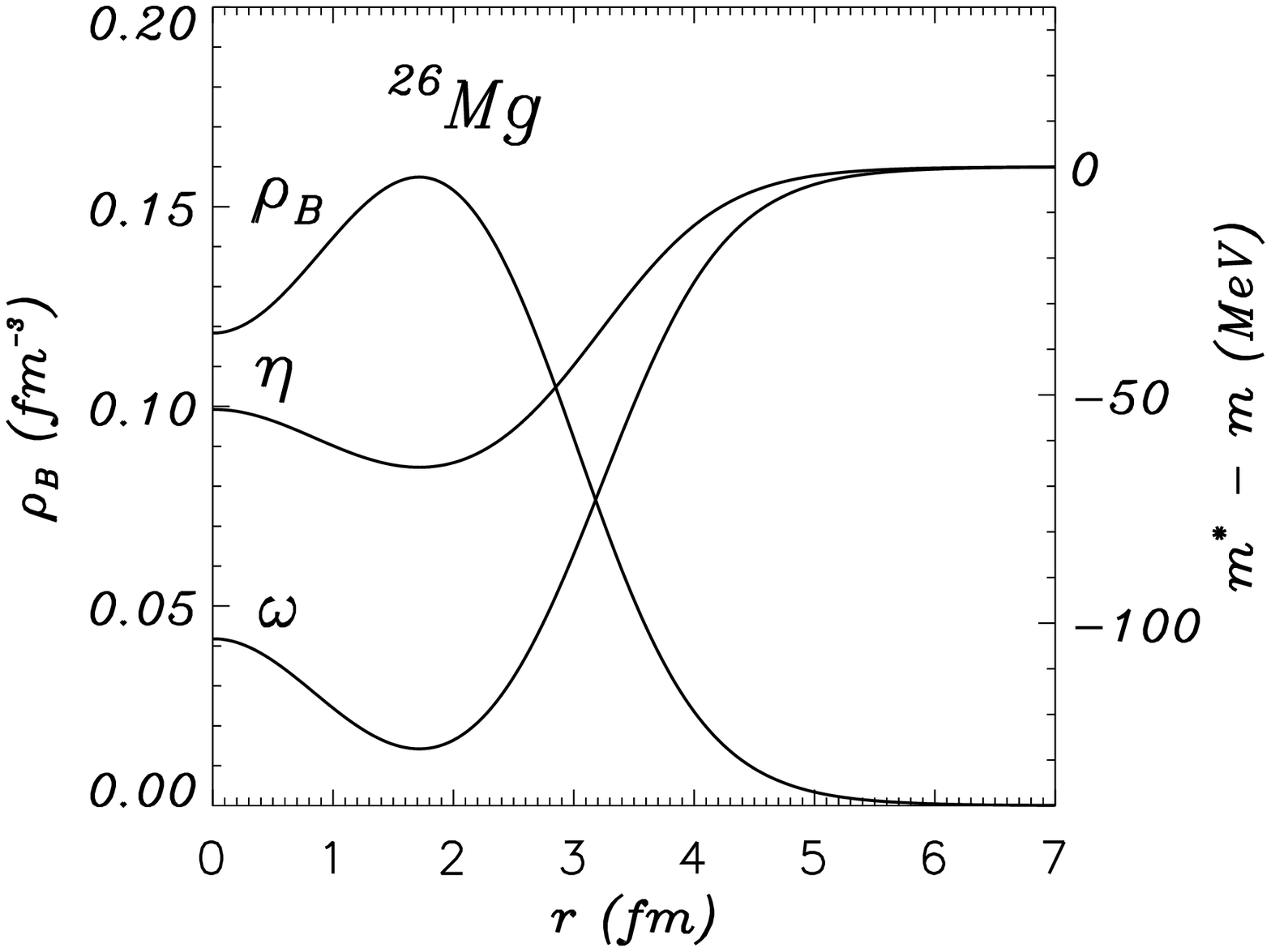}
\vspace{-2em}
\includegraphics[height=4cm,width=6cm]{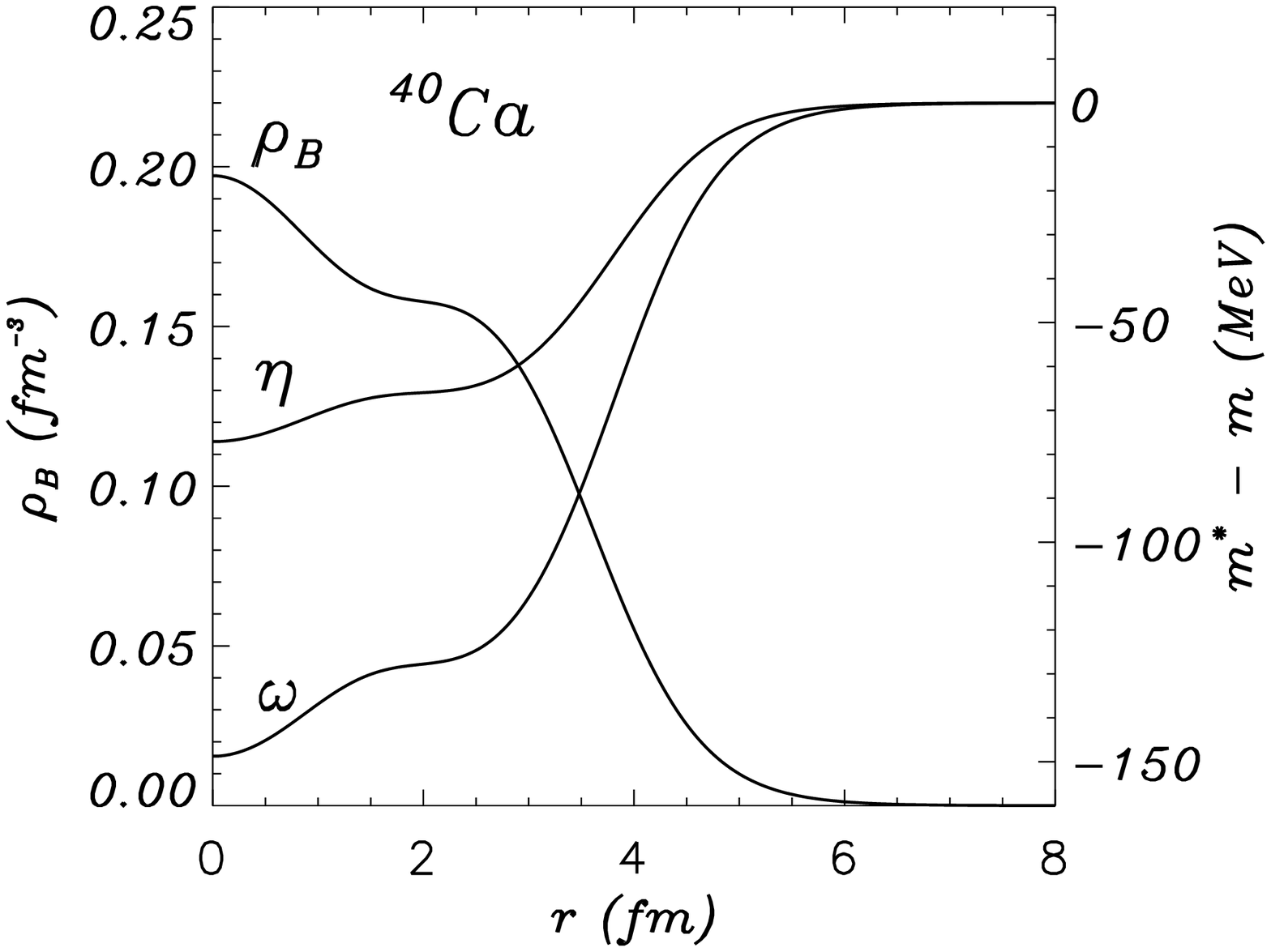}
\caption{Scalar potential for the $\omega$ and $\eta$ mesons 
in $^{26}$Mg and $^{40}$Ca.
\vspace{-1em}}
\label{etaopot}
\end{minipage}
\end{figure}

The normalized, static solution for the ground state quarks or antiquarks
with flavor $f$ in the hadron, $h$, may be written,  
$\psi_f (x) = N_f e^{- i \epsilon_f t / R_h^*}
\psi_f (\mbox{\boldmath $x$})$,
where $N_f$ and $\psi_f(\mbox{\boldmath $x$})$
are the normalization factor and
corresponding spin and spatial part of the wave function. The bag
radius in medium, $R_h^*$, 
will be determined through the
stability condition for the mass of the hadron against the
variation of the bag 
radius~\cite{Guichonf}
(see Eq.~(\ref{hmass})). The eigenenergies, $\epsilon_f$, in
the wave function in units of $1/R_h^*$, are given by
\begin{equation}
\left( \begin{array}{c}
\epsilon_u \\
\epsilon_{\bar{u}}
\end{array} \right)
= \Omega_q^* \pm R_h^* \left(
V^q_\omega
+ \frac{1}{2} V^q_\rho \right),\,\,
\left( \begin{array}{c} \epsilon_d \\
\epsilon_{\bar{d}}
\end{array} \right)
= \Omega_q^* \pm R_h^* \left(
V^q_\omega
- \frac{1}{2} V^q_\rho \right),\,\,
\epsilon_{s,c}
= \epsilon_{\bar{s},\bar{c}} =
\Omega_{s,c},
\label{cenergy}
\end{equation}
where $\Omega_q^*
{=}\sqrt{x_q^2{+}(R_h^* m_q^*)^2}$, with
$m_q^*{=}m_q{-}g^q_\sigma \sigma$ and
$\Omega_{s,c}{=}\sqrt{x_{s,c}^2{+}(R_h^* m_{s,c})^2}$.
The hadron masses
in a nuclear medium are calculated by
\begin{eqnarray}
m_h^* &=& \frac{(n_q + n_{\bar{q}}) \Omega_q^*
+ (n_{s,c} + n_{\bar{s},\bar{c}}) \Omega_{s,c} - z_h}{R_h^*}
+ {4\over 3}\pi R_h^{* 3} B,\qquad
\left. \frac{\partial m_h^*}
{\partial R_h}\right|_{R_h = R_h^*} = 0,
\label{hmass}
\end{eqnarray}
where $n_q$ ($n_{\bar{q}}$) and $n_{s,c}$ 
($n_{\bar{s},\bar{c}}$)
are the lowest mode light quark (antiquark) and strange, charm 
(antistrange, anticharm) quark numbers in the hadron, $h$, respectively,
and the $z_h$ parametrize the sum of the
center-of-mass and gluon fluctuation effects, and are assumed to be
independent of density. The parameters are determined in free space to
reproduce the corresponding masses.
We chose the values, $m_q$=5 MeV, $m_s$=250 MeV and  
$m_c$=1300 MeV for the current quark masses, and $R_N{=}0.8$
fm for the bag radius of the nucleon in free space. Other input
parameters and some of the quantities calculated are given in 
Refs.~\cite{Guichonf,Tsushimak,Tsushimaeta,Tsushimad}.
The quark-meson coupling constants, $g^q_\sigma$, $g^q_\omega$
and $g^q_\rho$, are adjusted to fit the nuclear 
saturation energy and density of symmetric nuclear matter, and the bulk
symmetry energy~\cite{Guichonf}. 
Exactly the same coupling constants, $g^q_\sigma$, $g^q_\omega$ and
$g^q_\rho$, are used for the light quarks in the mesons and hyperons as in 
the nucleon. However, in studies of the kaon system, we found that it was
phenomenologically necessary to increase the strength of the vector
coupling to the non-strange quarks in the $K^+$ (by a factor of
$1.4^2$) in order to reproduce the empirically extracted $K^+$-nucleus
interaction~\cite{Tsushimak}.  
We assume this also for the $D$ and $\bar{D}$ 
mesons~\cite{Tsushimad,Tsushimaj}.
The scalar ($U^{h}_s$) and vector ($U^{h}_v$) potentials 
felt by the hadrons, $h$,  
in nuclear matter are given by:
\begin{equation}
U_s = m^*_h - m_h,\quad
U_v =
  (n_q - n_{\bar{q}}) {V}^q_\omega - I_3 V^q_\rho, 
\qquad (V^q_\omega \to 1.4^2 {V}^q_\omega\,\, 
{\rm for}\, K,\bar{K},D,\bar{D}), 
\label{svpot}
\end{equation}
where $I_3$ is the third component of isospin projection  
of the hadron, $h$. 
We show in Figs.~\ref{dpot}~and~\ref{etaopot} some of the calculated 
mean field potentials. 

\section{$J/\Psi$ dissociation in nuclear matter~\cite{Tsushimaj}}

There is a great deal of interest in possible 
signals of Quark-Gluon Plasma (QGP)
formation (or precursors to its formation) and $J/\Psi$ suppression 
is a promising candidate.
On the other hand, there may be other mechanisms which produce an
increase in $J/\Psi$ absorption in a hot, dense medium. We are 
particularly interested in the rather exciting suggestion 
that the charmed mesons,
$D$, $\bar{D}$, $D^*$ and $\bar{D}^*$, should suffer substantial 
changes in their 
properties in a nuclear medium~\cite{Tsushimad,Tsushimaj}
(see Fig.~\ref{dpot}). This is expected to
have a considerable impact on charm production 
in heavy ion collisions~\cite{Tsushimaj}, 
although the mass of the $J/\Psi$ is expected to change 
by a tiny amount in nuclear matter within 
QMC~\cite{Guichonf,Tsushimad,Tsushimaj}.

The suppression of $J/\Psi$ production observed in relativistic 
heavy ion collisions, from $p{+}A$ up to central $S{+}U$ collisions,
has been relatively well understood. But recent data from
$Pb{+}Pb$ collisions shows a considerably stronger $J/\Psi$ 
suppression~\cite{Qm97,Vogt}. In an attempt to explain this 
``anomalous'' suppression,   
many authors have studied one of two possible
mechanisms, namely 
hadronic processes~\cite{Brodsky,Capella,Capella1,Martins,Mueller} 
and QGP formation~\cite{Matsui}.

In the hadronic dissociation scenario~\cite{Brodsky} 
the reactions involving the $J/\Psi$,
$\pi{+}J/\Psi {\to} D^*+\bar{D}, \bar{D}^*+D$ and
$\rho{+}J/\Psi {\to} D+\bar{D}$, are well known. 
The absorption of the $J/\Psi$ through these reactions have been 
found to be important in general and absolutely
necessary in order to fit the data on $J/\Psi$ production. 
(See Refs.~\cite{Cassing,Capella,Gavin} and references therein.)
$J/\Psi$ dissociation on comovers, combined with the absorption on 
nucleons, is the main mechanism proposed as an alternative to 
that of Matsui and Satz~\cite{Matsui}, i.e., the dissociation 
in a QGP. 

Within the hadronic scenario the crucial point is the required 
dissociation strength. In particular, one needs a total cross section for 
the $\pi,\rho{+}J/\Psi$ interaction of around $1.5~\sim~3$~mb
in order to explain
the data in heavy ion simulations~\cite{Cassing}. 
Recent calculations~\cite{Mueller} of the reactions,  
$\pi{+}J/\Psi {\to} D+\bar{D}^*, \bar{D}+D^*$ and 
$\rho{+}J/\Psi {\to} D+\bar{D}$, based on $D$ exchange, 
indicate a  much lower cross section than this. 
The main uncertainty in the discussion of the 
$J/\Psi$ dissociation on a meson gas is associated with the estimates 
of the $\pi,\rho{+}J/\Psi$ cross sections~\cite{Mueller}. 
According to the predictions for the $\pi{+}J/\Psi$ cross 
section~\cite{Mueller} taking into account the pion
kinetic energy and a thermal pion gas with average 
temperature of 150~MeV, one might conclude 
that the rate of this process is small~\cite{Tsushimaj} 
independent of the $\pi{+}J/\Psi$ dissociation
model used~\cite{Martins,Mueller}.

\begin{figure}[htb]
\begin{minipage}[htb]{80mm}
\vspace*{-2em}
\includegraphics[height=8cm,width=7cm]{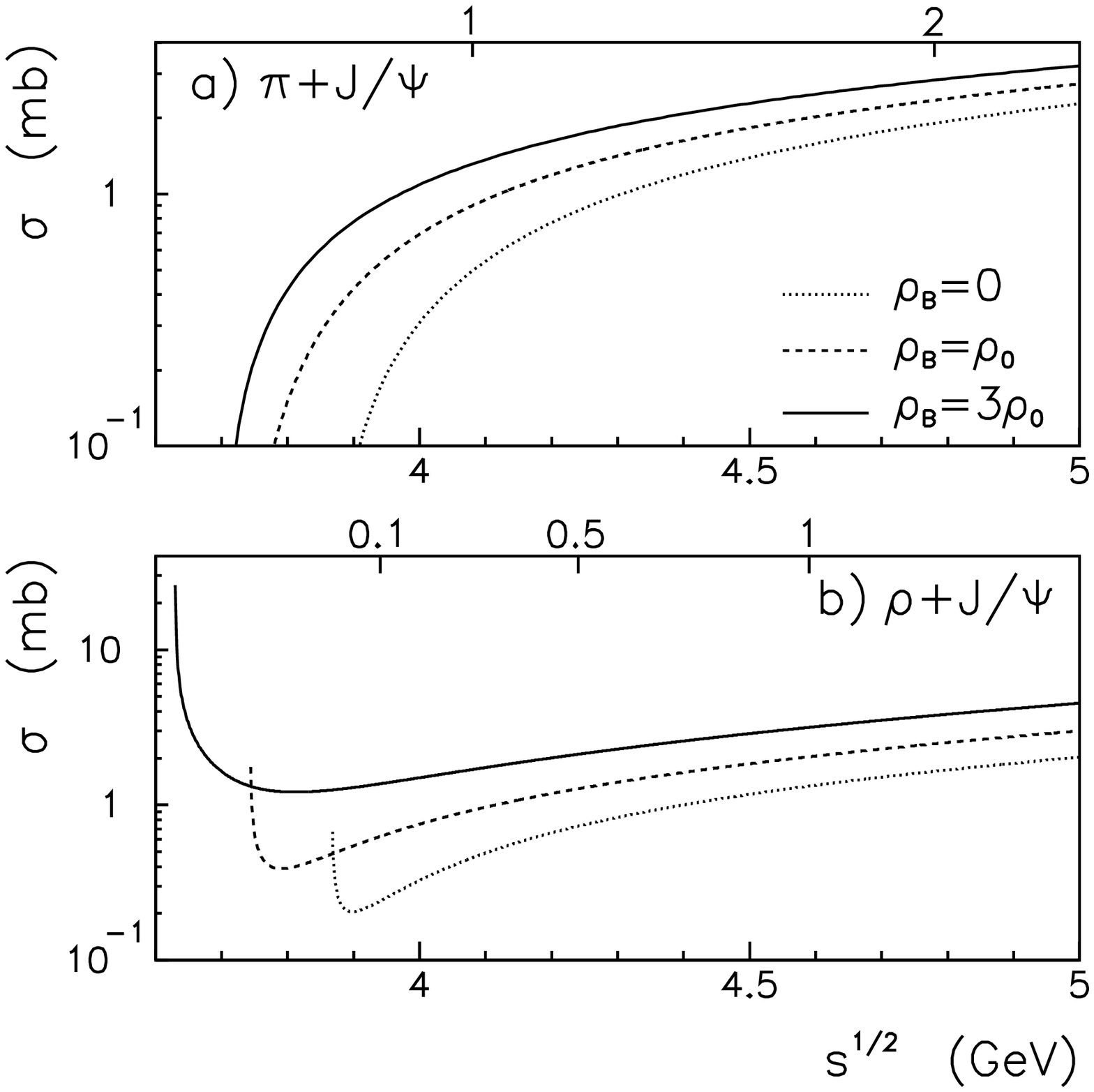}
\vspace{-3em}
\caption{$\pi{+}J/\Psi$ and $\rho{+}J/\Psi$ dissociation
cross sections as functions of the invariant collision energy, $s^{1/2}$.
Results are shown for vacuum (the dotted line)~\cite{Mueller},
$\rho_0$ (the dashed line)
and $3\rho_0$ (the solid line).
\vspace{-1em}}
\label{dcross}
\end{minipage}
\hspace{\fill}
\begin{minipage}[htb]{75mm}
\vspace*{-2em}
\includegraphics[height=8cm,width=7cm]{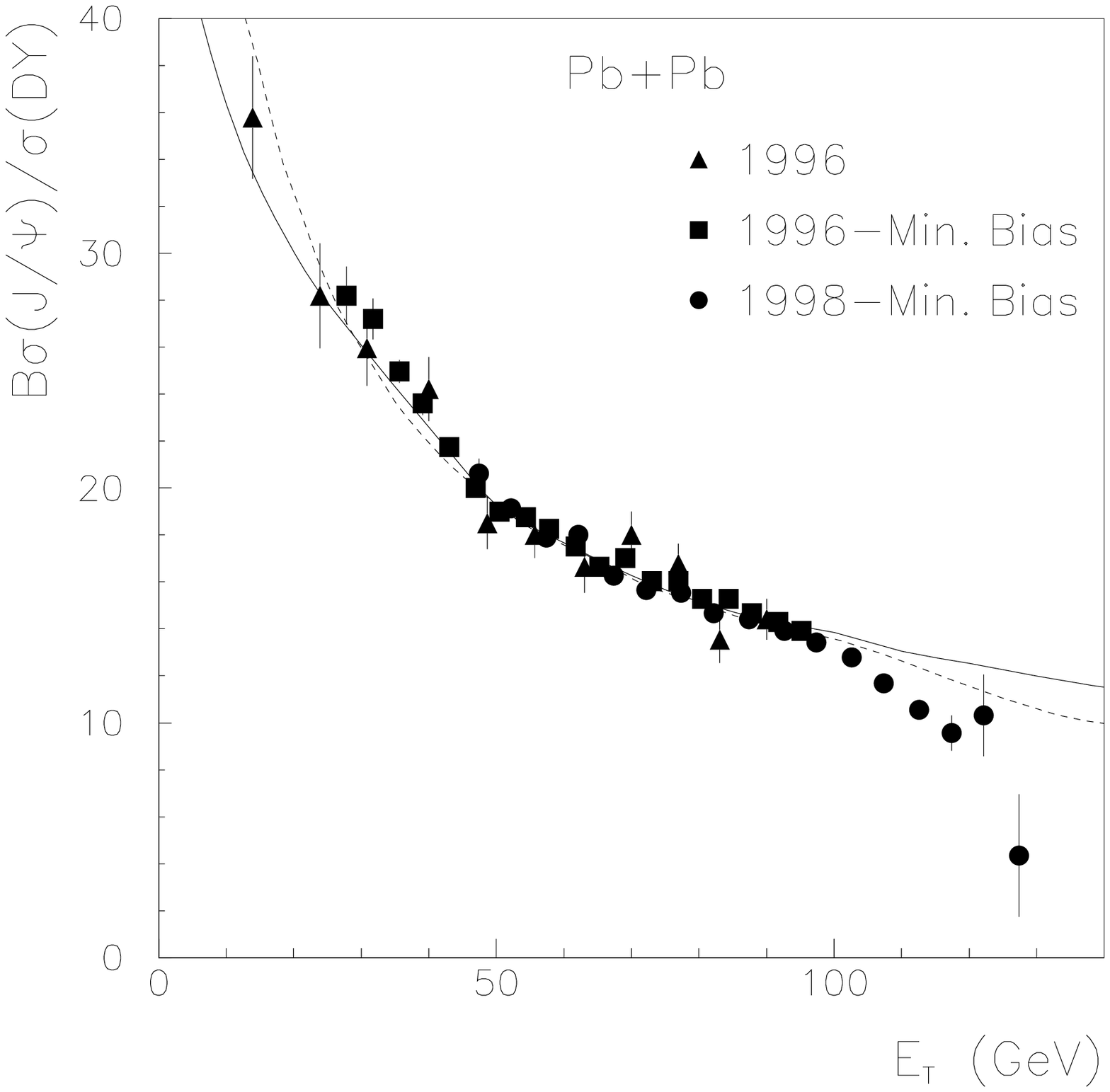}
\vspace{-3em}
\caption{The ratio of the $J/\Psi$ over Drell-Yan cross sections
from $Pb{+}Pb$ collisions as function of the transverse energy $E_T$.
Data are from Ref.~\cite{Qm97,Last}. The solid line shows our
calculations~\protect\cite{Tsushimaj}.
\vspace{-1em}}
\label{char7}
\end{minipage}
\end{figure}

However, this situation changes when
the in-medium potentials of the charmed mesons are taken into account,
because they lower the $\pi{+}J/\Psi{\to}\bar{D}{+}D^*$ and 
$\rho{+}J/\Psi{\to}\bar{D}{+}D^*$ reaction 
thresholds~\cite{Vogt,Cassing,Mueller} due to the effect of the vector and
scalar potentials felt by the charmed $D$, $D^*$ and $\rho$ mesons  
(see Fig.~\ref{dpot}). 
The cross sections calculated for the $\pi, \rho{+}J/\Psi$ collisions  
with the in-medium potentials are shown in Fig.~\ref{dcross}. 
Clearly the $J/\Psi$ absorption cross sections are substantially enhanced
for both the $\pi{+}J/\Psi$ and $\rho{+}J/\Psi$ reactions, not only 
because of the downward shift of the reaction threshold, but also 
because of the in-medium effect on the reaction amplitude. 
Moreover, now the $J/\Psi$ absorption
on comovers becomes both energy and density dependent -- a 
crucial finding given the situation in actual heavy ion collisions.
These effects have never been considered before. 
The absorption cross section has hitherto been taken as a constant. 

We found that the 
thermally averaged, in-medium $\pi{+}J/\Psi$ and $\rho{+}J/\Psi$
absorption cross sections,   
${\langle}{\sigma}v{\rangle}$, depend 
very strongly on the nuclear density~\cite{Tsushimaj}. Even for
$p_{J/\Psi}{=}0$, with a pion gas 
temperature of 120~MeV, which is close to the saturation pion density, 
the thermally averaged $J/\Psi$ absorption cross section on the 
pion at $\rho_B{=}3\rho_0$ is about a factor of 7 larger than that 
at $\rho_B{=}0$~\cite{Tsushimaj}.
As for the $\rho{+}J/\Psi$, the thermally averaged dissociation cross section  
at $\rho_B{=}3\rho_0$ becomes larger than 1~mb~\cite{Tsushimaj}.
Thus, the $J/\Psi$ absorption on $\rho$ mesons should be also appreciable, 
although it is expected the $\rho$ meson density is only 
half of the pion density in $Pb{+}Pb$ collisions~\cite{Cassing}.

In order to compare our results with the NA38/NA50 
data~\cite{Qm97,Last} on $J/\Psi$ suppression in $Pb{+}Pb$
collisions, we have adopted the heavy ion model proposed in
Ref.~\cite{Capella} with the $E_T$ model from 
Ref.~\cite{Capella1}. We  introduce the absorption
cross section on comovers as function of the density of comovers,
while the nuclear absorption cross section is taken 
as 4.5~mb~\cite{Capella1}. 
Our calculations are shown in Fig.~\ref{char7} by the solid line,
using the density dependent, thermally averaged 
cross section, ${\langle}{\sigma}v{\rangle}$, 
for $J/\Psi$ absorption on comovers~\cite{Tsushimaj}.
The dashed line in Fig.~\ref{char7} shows the calculations 
with the phenomenological constant cross section for $J/\Psi$ 
absorption on comovers, 
${\langle}{\sigma}v{\rangle}\simeq$1~mb given in Ref.~\cite{Capella1}.
Both curves clearly reproduce the data~\cite{Qm97} quite well, 
including most recent results from NA50 on the ratio
of $J/\Psi$ over Drell-Yan cross sections, as a function
of the transverse energy up to $E_T \simeq 100$~GeV. 
It is important to note that if one neglected the in-medium modification of 
the $J/\Psi$ absorption cross section the large cross section, 
${\langle}{\sigma}v{\rangle}$$\simeq$1~mb, could not be justified by 
microscopic theoretical calculations and thus
the NA50 data~\cite{Qm97,Last} could not be described. 
Furthermore, our calculations with in-medium modified absorption 
provide a significant improvement in the understanding of the 
data~\cite{Qm97} compared to the models quoted by NA50~\cite{Last}.

The basic difference between our results and those quoted by
NA50~\cite{Last} is that in previous 
heavy ion calculations~\cite{Vogt,Cassing,Capella} the cross section for 
$J/\Psi$ absorption on comovers was taken as a free parameter to
be adjusted to the data~\cite{Qm97,Last} and was never
motivated theoretically.

\section{\mb{\bm$\omega$-, $\eta$- and $\eta'$}-nuclear 
bound states~\cite{Tsushimaeta}} 

Next, we discuss the $\omega$, $\eta$-,
and $\eta'$-nuclear bound states~\cite{Tsushimaeta}.
We have solved the Klein-Gordon equation~\cite{Tsushimaeta}
using the calculated potentials (see Fig.~\ref{etaopot}):
\bg
&&\left[ \nabla^2 + E^{*2}_j - \tilde{m}^{*2}_j(r) \right]\,
\phi_j(\vr) = 0,\quad E^*_j \equiv E_j + m_j - i \Gamma_j/2,  
\qquad (j=\omega, \eta, \eta'),\\
&&\tilde{m}^*_j(r) =
m^*_j(r) - \frac{i}{2}
\left[ (m_j - m^*_j(r))
\gamma_j + \Gamma_j \right]
\equiv m^*_j(r) - \frac{i}{2} \Gamma^*_j (r),
\label{width}
\en
where $E^*_j$ is the complex valued, total energy  
of the meson, and we included
the widths of the mesons in a nucleus assuming a specific form
using $\gamma_j$, which are treated as phenomenological
parameters.
According to the estimates in Refs.~\cite{hayano,Friman},
the widths of the mesons in nuclei and at normal nuclear matter density
are $\Gamma^*_\eta \sim 30 - 70$ MeV~\cite{hayano}
and $\Gamma^*_\omega \sim 30 - 40$ MeV~\cite{Friman}, respectively.
Thus, we calculate the single-particle energies for the values
$\gamma_\omega = 0.2$, and $\gamma_\eta = 0.5$, which are
expected to correspond best with experiment, while
for the $\eta'$, $\Gamma^*_{\eta'} = 0$ is assumed.
For a comparison we give also the results for the $\omega$ 
calculated using the potential obtained in 
Quantum Hadrodynamics (QHD)~\cite{Saitoomega}.
%
\vspace{-1em}
\begin{table}[htb]
\begin{center}
\caption{
Calculated $\omega$-, $\eta$- and $\eta'$-nuclear bound
state energies (in MeV),
$E_j = Re (E^*_j - m_j)\,(j=\omega,\eta,\eta')$,
in QMC~\protect\cite{Tsushimaeta} and those for the $\omega$ in
QHD with $\sigma$-$\omega$ mixing effect~\protect\cite{Saitoomega}.
The complex eigenenergies are given by,
$E^*_j = E_j + m_j - i \Gamma_j/2$.
\qquad (* not calculated)
}
\begin{tabular}[t]{lc|cc||c||cc|cc}
\hline \hline
& &$\bm \gamma_\eta=0.5$ &(QMC) &(QMC) &$\bm \gamma_\omega$=0.2
&(QMC) &$\bm \gamma_\omega=0.2$ &(QHD)\\
\hline \hline
& &$E_\eta$ &$\Gamma_\eta$ &$E_{\eta'}$ &$E_\omega$ &$\Gamma_\omega$
&$E_\omega$ &$\Gamma_\omega$\\
\hline
$^{6}_j$He &1s &-10.7&14.5 & * &-55.6&24.7 &-97.4&33.5 \\
\hline
$^{11}_j$B &1s &-24.5&22.8 & * &-80.8&28.8 &-129&38.5 \\
\hline
$^{26}_j$Mg &1s &-38.8&28.5 & * &-99.7&31.1 &-144&39.8 \\
            &1p &-17.8&23.1 & * &-78.5&29.4 &-121&37.8 \\
            &2s & --- & --- & * &-42.8&24.8 &-80.7&33.2  \\
\hline \hline
$^{16}_j$O &1s &-32.6&26.7 &-41.3 &-93.4&30.6 &-134&38.7 \\
           &1p &-7.72&18.3 &-22.8 &-64.7&27.8 &-103&35.5 \\
\hline
$^{40}_j$Ca &1s &-46.0&31.7 &-51.8 &-111&33.1  &-148&40.1 \\
            &1p &-26.8&26.8 &-38.5 &-90.8&31.0 &-129&38.3 \\
            &2s &-4.61&17.7 &-21.9 &-65.5&28.9 &-99.8&35.6  \\
\hline
$^{90}_j$Zr &1s &-52.9&33.2 &-56.0 &-117&33.4  &-154&40.6 \\
            &1p &-40.0&30.5 &-47.7 &-105&32.3  &-143&39.8 \\
            &2s &-21.7&26.1 &-35.4 &-86.4&30.7 &-123&38.0 \\
\hline
$^{208}_j$Pb &1s &-56.3&33.2 &-57.5 &-118&33.1 &-157&40.8 \\
             &1p &-48.3&31.8 &-52.6 &-111&32.5 &-151&40.5 \\
             &2s &-35.9&29.6 &-44.9 &-100&31.7 &-139&39.5 \\
\hline \hline
\end{tabular}
\end{center}
\end{table}
\vspace{-1em}

Our results suggest that one should expect to
find bound $\omega$-, $\eta$- and $\eta'$-nuclear states 
for all nuclei investigated and relatively wide range of 
the in-medium meson widths~\cite{Tsushimaeta}.
(For the predictions made by the other approaches, 
see Refs.~\cite{hayano,Friman}.)  

\section{Summary}

We have presented two possible signals for a change of 
meson properties in a nuclear medium.  
First, we discussed the impact on the  
observed $J/\Psi$ suppression in relativistic heavy ion collisions, 
which has received a lot of interest recently because of 
the Quark-Gluon Plasma. 
Second, we discussed the prediction of 
$\omega$-, $\eta$- and $\eta'$-nuclear bound states.   
Both of these signals, for which many experiments are currently 
being perfomed or 
planned, certainly will provide important 
information about the in-medium properties of hadrons.

%

%

\begin{thebibliography}{99}
\bibitem{Guichon}
        P.A.M. Guichon, Phys. Lett. {\bf B200} (1989) 235.
\bibitem{Guichonf}
        P.A.M. Guichon et al., Nucl. Phys. {\bf A601} (1996) 349;
        K. Saito et al., Nucl. Phys. {\bf A609} (1996) 339;
        K. Tsushima et al., Nucl. Phys. {\bf A630} (1998) 691.
\bibitem{Tsushimak}
        K. Tsushima et al., Phys. Lett. {\bf B429} (1998) 239; 
        (E) ibid. {\bf B436} (1998) 453.
\bibitem{Tsushimaeta}
        K. Tsushima, D.H. Lu, A.W. Thomas, K. Saito,
        Phys. Lett. {\bf B443} (1998) 26.
\bibitem{Tsushimad}
        K. Tsushima et al., Phys. Rev. {\bf C59} (1999) 2824.
\bibitem{Tsushimaj}
        A. Sibirtsev, K. Tsushima, K. Saito, A.W. Thomas,
        nucl-th/9904015, to app. in {\bf PLB}.
\bibitem{Saitoomega}
        K. Saito, K. Tsushima, D.H. Lu, A.W. Thomas,
        Phys. Rev. {\bf C59} (1999) 1203.
\bibitem{Qm97} Quark Matter '97, Nucl. Phys. {\bf A638} (1998);
        M. C. Abreu et al. (NA50 Collaboration),
        Phys. Lett. {\bf B410}  (1997) 337; 
        Phys. Lett. {\bf B450} (1999) 456.
\bibitem{Vogt}
        R. Vogt, Phys. Rep. {\bf 310} (1999) 197. 
\bibitem{Cassing}
        W. Cassing and E. Bratkovskaya, Phys. Rep. {\bf 308} (1999) 65.
\bibitem{Brodsky}
        S.J. Brodsky and A.H. Mueller, Phys. Lett. {\bf B206} (1988) 685;
        S. Gavin, M. Gyulassy and A. Jackson, 
        Phys. Lett. {\bf B207} (1988) 257; 
        R. Vogt et al., Phys. Lett. {\bf B207} (1988) 264;
        J.-P. Blaizot and J.-Y. Ollitrault, 
        Phys. Rev. {\bf D39} (1989) 232;
\bibitem{Capella}
        N. Armesto and  A. Capella, J. Phys. {\bf G23} (1997) 1969; 
        Phys. Lett. {\bf B430} (1998) 23; 
        N. Armesto, A. Capella and E.G. Ferreiro, 
        Phys. Rev. {\bf C59} (1999) 395. 
\bibitem{Capella1}
        A. Capella, E.G. Ferreiro and A.B. Kaidalov,
        hep-ph/0002300.
\bibitem{Martins}
        K. Martins, D. Blaschke and E. Quack,
        Phys. Rev. {\bf C51} (1995) 2723;
        D. Kharzeev, H. Satz, A. Syamtomov, G. Zinovev,
        Phys. Lett. {\bf B389} (1996) 595.
\bibitem{Mueller}
        S.G. Matinyan and B. M\"uller,
        Phys. Rev. {\bf C58}  (1998) 2994;
        B. M\"uller, Nucl. Phys. {\bf A661} (1999) 272.
\bibitem{Matsui}
        T. Matsui and H. Satz, Phys. Lett. {\bf B178} (1986) 416. 
\bibitem{Gavin}
        S. Gavin and R. Vogt, 
        Nucl. Phys. {\bf B345} (1990) 104;
        S. Gavin, H. Satz, R.L. Thews and R. Vogt,
        Z. Phys. {\bf C61} (1994) 351.
\bibitem{Last}
        M. C. Abreu et al. (NA50 Collaboration),
        Phys. Lett. {\bf B477} (2000) 28.
%
\bibitem{hayano}R.S. Hayano et al., proposal for
GSI/SIS, September, 1997;
R.S. Hayano, S. Hirenzaki and A. Gillitzer, Eur. Phys. J. {\bf A6} (1999) 99.
%
\bibitem{Friman}B. Friman, nucl-th/9801053;
F. Klingl and W. Weise, hep-ph/9802211; 
F. Klingl, T. Waas, W. Weise, Nucl. Phys. {\bf A650} (1999) 299.
%
\end{thebibliography}
\end{document}